\documentclass[sigconf]{acmart}

\usepackage{multirow} 
\usepackage{array}
\usepackage{makecell}
\usepackage{subcaption}
\usepackage{enumitem}

\AtBeginDocument{%
  \providecommand\BibTeX{{%
    \normalfont B\kern-0.5em{\scshape i\kern-0.25em b}\kern-0.8em\TeX}}}

\setcopyright{acmcopyright}
\copyrightyear{2025}
\acmYear{2025}
\setcopyright{acmlicensed}\acmConference[RecSys '25]{Proceedings of the Nineteenth ACM Conference on Recommender Systems}{September 22--26, 2025}{Prague, Czech Republic}
\acmBooktitle{Proceedings of the Nineteenth ACM Conference on Recommender Systems (RecSys '25), September 22--26, 2025, Prague, Czech Republic}
\acmDOI{10.1145/3705328.3748036}
\acmISBN{979-8-4007-1364-4/2025/09}

\begin{document}


\title{SGCL: Unifying Self-Supervised and Supervised Learning \\ for Graph Recommendation}








\author{Weizhi Zhang$^{1}$, Liangwei Yang$^{2\dagger}$, Zihe Song$^1$, Henrry Peng Zou$^1$, \\ Ke Xu$^1$, Yuanjie Zhu$^1$, Philip S. Yu$^1$}
\affiliation{%
\institution{$^1$University of Illinois Chicago
\city{Chicago}
\country{USA}}
}
\affiliation{%
\institution{$^2$Salesforce AI Research
\city{Palo Alto}
\country{USA}}
}

\thanks{$^\dagger$Corresponding author: liangwei.yang@salesforce.com}

\renewcommand{\shortauthors}{Weizhi Zhang et al.}

\begin{abstract}
Recommender systems (RecSys) are essential for online platforms, providing personalized suggestions to users within a vast sea of information. Self-supervised graph learning seeks to harness high-order collaborative filtering signals through unsupervised augmentation on the user-item bipartite graph, primarily leveraging a multi-task learning framework that includes both supervised recommendation loss and self-supervised contrastive loss. However, this separate design introduces additional graph convolution processes and creates inconsistencies in gradient directions due to disparate losses, resulting in prolonged training times and sub-optimal performance. In this study, we introduce a unified framework of Supervised Graph Contrastive Learning for recommendation (SGCL) to address these issues. SGCL uniquely combines the training of recommendation and unsupervised contrastive losses into a cohesive supervised contrastive learning loss, aligning both tasks within a single optimization direction for exceptionally fast training. Extensive experiments on three real-world datasets show that SGCL outperforms state-of-the-art methods, achieving superior accuracy and efficiency. 
\end{abstract}



\maketitle

\section{Introduction}

\begin{figure}[htpb]
    \centering
    \includegraphics[width=0.95\linewidth]{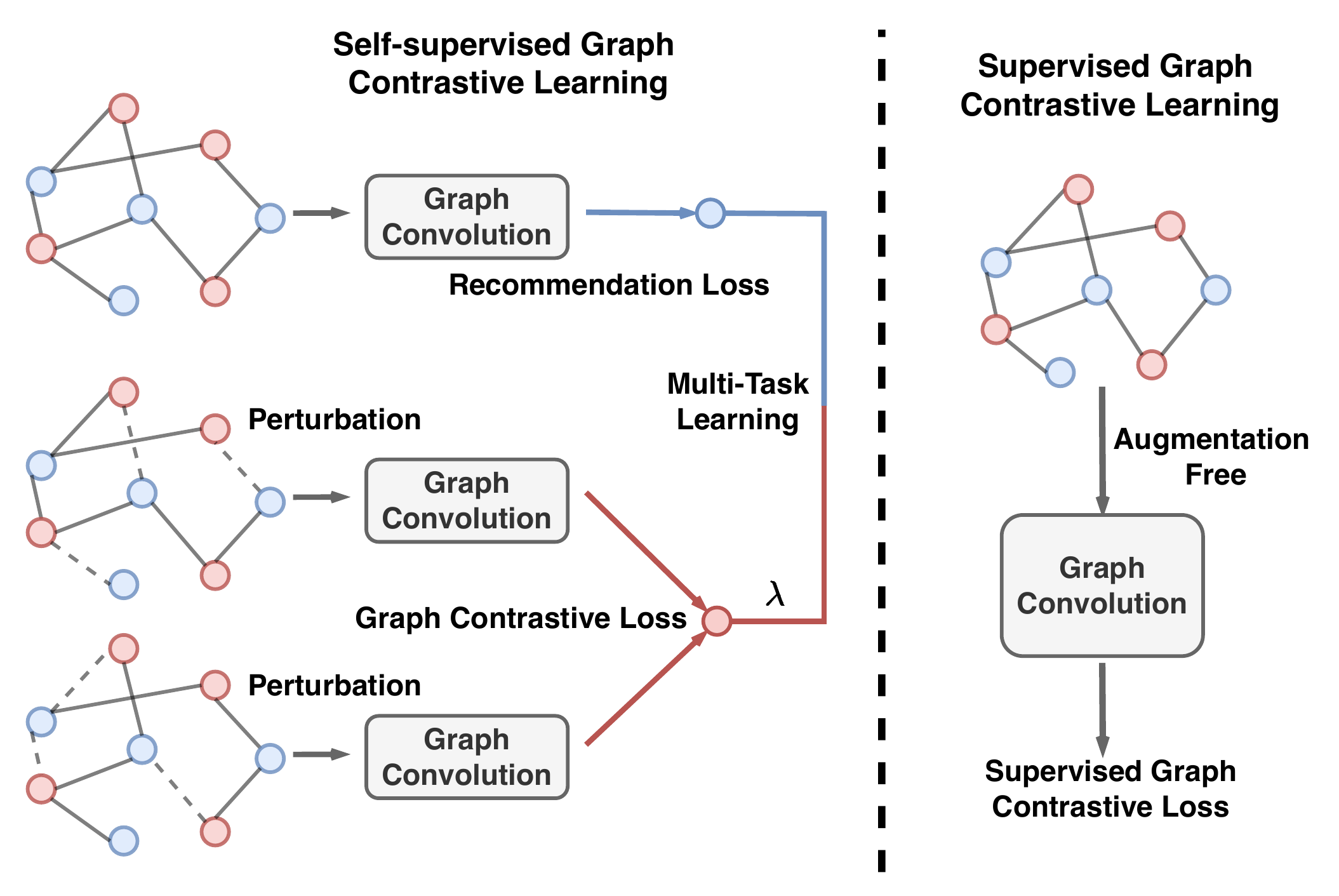}
    \caption{The left figure illustrates the pipeline of traditional SSL RecSys, including graph augmentation, three rounds of graph convolutions, and two distinct loss optimization controlled by $\lambda$.
    The right figure depicts our SGCL framework, consisting of a single graph encoder and a unified SGCL loss.}
    \label{fig: intro}
\end{figure}

Recommender systems (RecSys) have emerged as a cornerstone in the architecture of modern online systems and web applications, playing a pivotal role in filtering and personalizing the information presented to users~\cite{zhang2025cold, yang2024item}. These systems are engineered to deliver tailor-made item recommendations that align with individual user preferences~\cite{zhang2025personaagent}. The scope of RecSys extends across various domains, including but not limited to, digital retailing~\cite{wang2020time,hwangbo2018recommendation,zhang2023dual}, social networking platforms~\cite{jamali2010matrix,fan2019graph}, and video-sharing services~\cite{wei2023multi}. Among the methodologies employed in RecSys, collaborative filtering (CF) stands out as a prominent approach~\cite{he2017neural, xu2023graph, xu2025graph}. 
CF algorithms function by extrapolating patterns from historical user-item interactions to identify and recommend items to users with similar interaction profiles. However, this approach is inherently limited by simple supervised recommendation losses and sparse user-item interactions to model and predict user preferences accurately~\cite{zhang2023dual, zhang2025cold, yang2025cold, zhang2025llminit}.

Graph contrastive learning has been increasingly integrated into RecSys as a solution to mitigate these challenges. This approach primarily involves data augmentation on bipartite graph structure, where high-order structural signals are extracted via graph neural networks~\cite{he2020lightgcn,wang2019neural,zhang2024we, ma2023graph}. Concurrently, graph contrastive learning establishes a self-supervised learning (SSL) task, aimed at enhancing representation expressiveness by distinguishing the nodes in the bipartite graph. This is achieved through the creation of varied representations from different views, derived from graph augmentation techniques or embedding perturbations. Existing methodologies in this domain~\cite{wu2021self,lin2022improving,cai2023lightgcl,yu2022graph, yang2023generative} typically employ a decoupled design approach. This involves the utilization of a recommendation-specific supervised loss function~\cite{rendle2012bpr}, designed to learn from user-item interaction signals, in conjunction with a distinct contrastive learning loss~\cite{oord2018representation}, which focuses on assimilating knowledge from self-supervised paradigm.

Despite their enhanced performance over traditional models that solely consider original user-item interactions~\cite{ying2018graph, wang2019neural, he2020lightgcn}, current graph contrastive learning methods in RecSys still confront \textbf{\textit{Ineffective Optimization}} challenges that impede further improvements in performance and efficiency.
It arises from the prevalent practice of employing a decoupled design for recommendation supervision and self-supervised contrastive signals as shown in Figure~\ref{fig: intro}. This approach necessitates the redundant graph encoders to perform the graph convolutions on the whole bipartite graph, posing extreme challenges on the large-scale graph recommendation. Beyond that, the development of two distinct loss functions, while aiming to learn from different tasks, optimize the same set of parameters, leading to potential conflicts and inconsistent gradients. Such inconsistent gradients have been identified as a primary factor contributing to unstable multi-task training and performance degradation, as highlighted in recent studies~\cite{liu2023deep, tang2023improving}. This not only expends considerable computational resources but also hampers further enhancement and training efficiency of the model.

In this paper, we introduce Supervised Graph Contrastive Learning for Recommendation (SGCL), a novel framework designed to address the identified obstacles. To tackle the \textbf{\textit{ineffective optimization}}, SGCL incorporates a unique supervised graph contrastive learning loss. This loss function integrates the user-item interaction supervision signal directly within the self-supervised loss, thereby merging supervised and self-supervised contrastive learning tasks into a single optimization objective. Such an approach not only addresses unstable and inconsistent optimization inherent in the multi-task learning framework but also obviates the necessity of redundant graph convolutions and extensive hyperparameter tuning of weight $\lambda$ as illustrated in Figure~\ref{fig: intro}. 
In addition, since SGCL loss computations eliminate the negative sampling process and data augmentations, the overall training is extremely time-efficient.
To promote future research in supervised graph contrastive learning RecSys, we have made it available as an open-source resource \textcolor{blue}{\url{https://github.com/DavidZWZ/SGCL}}. 
Our contributions are summarized as follows:

\begin{itemize}[leftmargin=*]
    \item Design of efficient framework: We have identified the \textbf{\textit{Ineffective Optimization}} as a key bottleneck and designed a streamlined framework to eliminate redundant graph convolutions and complex data augmentations, marking an extremely efficient pipeline. 
    \item Development of a novel loss: Methodologically, we have developed an innovative supervised graph contrastive learning loss, which uniquely integrates training of supervised and self-supervised tasks within a singular objective function. 
    \item Experimentally, we conduct extensive experiments on three real-world datasets to test the effectiveness of SGCL. It achieves the highest recommendation score with remarkably low training time, demonstrating the exceptional efficiency of SGCL.
\end{itemize}

\section{Preliminaries}

\subsection{Self-supervised Graph Contrastive Learning}
Recent approaches \cite{wu2021self,lin2022improving,yu2022graph,cai2023lightgcl, yang2023generative} for self-supervised graph recommendation all rely on the multi-task learning framework and jointly learn from contrastive learning tasks and supervised recommendation tasks on the node representations as shown in Figure~\ref{fig: intro}.

In a formal definition, the joint learning scheme in self-supervised graph recommendation is to optimize the SSLRec loss function:
\begin{equation}
    \mathcal{L}_{\text {sslrec}}=\mathcal{L}_{\text {rec }}+\lambda \mathcal{L}_{gcl},
\end{equation}
where the $\mathcal{L}_{\text {sslrec}}$ consists of the traditional recommendation loss $\mathcal{L}_{\text {rec }}$, such as BPR \cite{rendle2012bpr}, and the contrastive learning (CL) loss $\mathcal{L}_{gcl}$, with $\lambda$ to weight the two losses. As the inputs are the bipartite user-item interaction graph, the contrastive learning process is conducted concurrently for both the user and item side. Based on the InfoNCE \cite{oord2018representation}, the user-side CL loss $\mathcal{L}^u_{gcl}$ is formulated as:
\begin{equation}\label{eq: sgl}
    \mathcal{L}^u_{gcl}=\sum_{i \in \mathcal{B}_u}-\log \frac{ \exp \left(u_i^{\prime \top} u_i^{\prime \prime} / \tau\right)}{\sum_{k \in \mathcal{B}_u} \exp\left(u_i^{\prime \top} u_k^{\prime \prime} / \tau\right)},
\end{equation}
where $i, k$ are the nodes (users) in the user batch $\mathcal{B}_u$ and $u_i^{\prime}, u_i^{\prime \prime}$ are two views of node representation after the data augmentations. Note that the $\tau$ is a hyperparameter that controls the temperature of the InfoNCE loss and applies the penalties on hard negative samples. Then, the item-side CL loss $\mathcal{L}^v_{gcl}$ can be calculated in the same format. Finally, the CL loss will be the summation of both user and item sides $\mathcal{L}_{gcl} = \mathcal{L}^u_{gcl}+\mathcal{L}^v_{gcl}$.
In general, the graph contrastive learning loss $\mathcal{L}_{gcl}$ encourages the similarity between two variant embeddings $u_i^{\prime}$ and $u_i^{\prime \prime}$ from the same node while maximizing the inconsistency among all the other nodes in batch.

\section{Proposed solutions}

{In this section, we introduce the proposed Supervised Graph Contrastive Learning (SGCL) in detail. Firstly, we elaborate on the novel Supervised Graph Contrastive Loss that integrates the training of both supervised recommendation and contrastive learning losses. Then, a detailed time complexity investigation is conducted to demonstrate the efficiency of the proposed SGCL.}

\subsection{Supervised Graph Contrastive Learning}
In the context of graph-based recommendation tasks, suppose we observe a pair of interacted users and items with corresponding initial input ID embeddings $u^0_i$ and $v^0_j$, and we adopt the light graph convolution as in \cite{he2020lightgcn}, by consistently aggregating the connected neighbors' representations:
\begin{equation}
\begin{aligned}
u^{k+1}_i = \sum_{j\in N_i} \frac{1}{\sqrt{|N_i|} \sqrt{|N_j|}}v^{k}_j,\\
v^{k+1}_j = \sum_{i\in N_j} \frac{1}{\sqrt{|N_i|} \sqrt{|N_j|}}u^{k}_i,\\
\end{aligned}
\end{equation}
where $u^{k}_i$ and $v^{k}_j$ are embeddings of user $u_i$ and item $v_j$ at $k$-th layer, respectively. The normalization employs the average degree $\frac{1}{\sqrt{|N_i|} \sqrt{|N_j|}}$ to temper the magnitude of popular nodes after graph convolution in each layer.
Afterward, the collaborative filtering final embedding is to synthesize the layer-wise representations:
\begin{equation}
\begin{aligned}
u_i = \sum^{K}_{k=0} \alpha_k u^k_i; \quad
v_j = \sum^{K}_{k=0} \alpha_k v^k_j,
\end{aligned}
\end{equation}
where $\alpha_k$ is the representation weight of the $k$-th layer and $K$ is total numbers of the layers. Note that we assume $\alpha_k = \frac{1}{K+1}$ unless mentioned in the following sections.

For the self-supervised graph recommendation, the contrastive loss $\mathcal{L}_{gcl}$ in Equation~\ref{eq: sgl} is incapable of explicitly learning supervised signals from existing interactions within graph data. Therefore, recommendation loss $\mathcal{L}_{rec}$ is required for label utilization, whilst an extra weight $\lambda$ should be dedicatedly tuned to balance and mitigate the gradient inconsistency. Towards uniting the power of the supervised signal and self-supervised learning strategy in the graph recommendations, we devised a new type of supervised graph contrastive learning (SGCL) loss as follows:
\begin{equation}
\mathcal{L}_{sgcl}=\sum_{(i, j) \in \mathcal{B}}-\log \frac{\exp \left({u}_i^{\top} {v}_j / \tau\right)}{
\sum_{(i^{\prime},j^{\prime}) \in \mathcal{B}} \left(\exp ( {u}_i^{\top}  {u}_{i^{\prime}}  / \tau) + \exp ( {v}_j^{\top}  {v}_{j^{\prime}}  / \tau)\right)}, 
\end{equation}
where $(i, j)$ are paired user-item interaction in batch $\mathcal{B}$ and $(i^\prime, j^\prime)$ are the rest user-item representations in batch. It is noted that this version of the loss function does not involve any type of data augmentation, and the user-item interactions are considered positive pairs. Therefore, both complex graph augmentation and time-consuming negative sampling processes are avoided, making the SGCL training process extremely efficient. More importantly, we alleviate the need to use multi-task learning for model optimization and use only one supervised graph contrastive loss during training.

\subsection{Time Complexity}
Here, we conduct the time complexity analysis on SGCL and compare it with predominant baseline methods LightGCN~\cite{he2020lightgcn} and SGL \cite{wu2021self} in graph recommendation. We first denote $|\mathcal{E}|$ as the number of edges. Then, let $K$ be the number of layers in graph convolution, and ${d}$ be the embedding size. $p$ represents the probability of retaining the edges in SGL and $N_{v}$ indicates the average number of item candidates in negative sampling.

Then, we can derive that: 1) For the normalization of the adjacency matrix, both LightGCN and SGCL do not contain the graph structure manipulation while SGL drops the edges in the graph data to create two versions of the modified adjacency matrix, and thus the time complexity is approximately three times of our methods. 
2) In the graph convolution, as SGL changed the graph structure twice in contrastive learning, it requests an extra $4p|\mathcal{E}|Kd$ for graph convolution. Whereas no augmentations are implemented before graph convolution in LightGCN and SGCL, resulting in more efficient $2|\mathcal{E}|Kd$ time consumption. 
3) While SGLC obviates the necessity of BPR loss computation, LightGCN and SGL still need $2BN_v$ for negative sampling and $2Bd$ for calculating the 2B pairs of interactions, respectively.
4) Regarding the graph contrastive loss, SGL separates the user and item loss calculation into $\mathcal{L}^u_{gcl}$ and $\mathcal{L}^v_{gcl}$, and thus the total time cost is $O(2Bd + 2B^2d)$. Since SGCL is based on a joint embedding learning process of the users and items, the time complexity in the batch is $O(Bd + 2B^2d)$. It is noted that, as all experiments are conducted on GPU parallel computations, therefore the time spent in loss computation differs slightly.

\begin{table}[ht]
\centering
\caption{Time complexity comparison of LightGCN, SGL, and SGCL in different steps of graph recommendation.}
\small
\begin{tabular}{l|m{1.6cm}|m{2.3cm}|m{1.8cm}}
\toprule
 Steps & LightGCN & SGL & SGCL\\
\midrule
\makecell[l]{Adj.\\Matrix} &  \( O(2|\mathcal{E}|) \) & \( O((2 + 4p)|\mathcal{E}|) \) & \( O(2|\mathcal{E}|) \)\\
\midrule
\makecell[l]{Graph \\ Conv.} & \( O(2|\mathcal{E}|Kd) \) & \( O((2 + 4p)|\mathcal{E}|Kd) \)   & \( O(2|\mathcal{E}|Kd) \)\\
\midrule
Sampling & \( O(BN_v) \) & \( O(BN_v) \) & -\\
\midrule
BPR Loss  & \( O(2Bd) \) & \( O(2Bd) \) & -\\
\midrule
GCL Loss  & -- & \( O(2Bd + 2B^2d) \) & \( O(Bd + 2B^2d) \) \\
\bottomrule
\end{tabular}
\label{tab:complexity}
\end{table}

\section{Evaluation}

\subsection{Experimental Setup}

To evaluate the superior performance and efficiency of the proposed SGCL, we conduct the experiments on two public real-world datasets from Amazon~\footnote{\url{https://jmcauley.ucsd.edu/data/amazon/links.html}}: Beauty and Toys-and-Games, varying in domains and scales.
We use the 5-core filtering setting to ensure the data quality for evaluation and split all datasets into training, validation, and testing with the ratio (8:1:1).  
In baseline selection, we compares MF methods (BiasMF~\cite{koren2009matrix}, NeuMF~\cite{he2017neural}), graph models (NGCF~\cite{wang2019neural}, LightGCN~\cite{he2020lightgcn}), self-supervised graph approaches (NCL~\cite{lin2022improving}, SGL~\cite{wu2021self}, LightGCL~\cite{cai2023lightgcl}, SimGCL~\cite{yu2022graph}). 
For the training of all the baselines, we carefully search their hyperparameters for different datasets to ensure a fair comparison. For all approaches, batch size $B$ and embedding size $d$ are set to 1,024 and 64. In SGCL, the temperature $\tau$ is set to 0.2, as it was found to be optimal.

\begin{table}[htpb]
\small
\setlength{\tabcolsep}{2pt} 
\begin{center}
\caption{Performance comparison on Beauty and Toys-and-Games in terms of NDCG and Recall.}
  \label{tab:main}
  \begin{tabular}{ccccccccc}
    \toprule
    \multirow{2}{*}{Method} & \multicolumn{4}{c}{Beauty} & \multicolumn{4}{c}{Toys-and-Games}\\
    
    \cmidrule(r){2-5} \cmidrule(r){6-9} 
    & R@20 & N@20 & R@50 & N@50 
    & R@20 & N@20 & R@50 & N@50  \\
    
    \midrule
    BiasMF & 0.1152 & 0.0544 & 0.1754 & 0.0668 & 0.0981 & 0.0479 & 0.1624 & 0.0625  \\
    NeuMF & 0.0720 & 0.0320 & 0.1317 & 0.0490 & 0.0690 & 0.0345 & 0.1094 & 0.0428 \\
    \midrule
    
    NGCF & 0.1126 & 0.0517 & 0.1699 & 0.0625 & 0.0944 & 0.0429 &  0.1532 & 0.0561 \\
    LightGCN & 0.1224  & 0.0583 & 0.1903 & 0.0718 & 0.1159 & 0.0565 & 0.1759 & 0.068 \\

    \midrule
    NCL & 0.1264 & 0.0613 & 0.1948 & 0.0721 & 0.117 & 0.0566 & 0.1782 & 0.0686 \\
    SGL & {0.1335} & {0.0634} & {0.2048} & {0.078} & \underline{0.1248} & \underline{0.0595} & \underline{0.1869} & \underline{0.072} \\
    LightGCL & 0.1278 & 0.0596 & 0.1905 & 0.0697 & 0.1177 & 0.0549 & 0.1772 & 0.0647 \\
    SimGCL & \underline{0.1345}	& \underline{0.064} & \underline{0.2058} & \underline{0.0799} & 0.1162 & 0.0593 & 0.1738 & 0.0713 \\
    \midrule
    
    SGCL & \textbf{0.141} & \textbf{0.0677} &  \textbf{0.2105} & \textbf{0.0807} & \textbf{0.1317} & \textbf{0.0626} & \textbf{0.1965} & \textbf{0.0774}  \\
  \bottomrule
\end{tabular}
\end{center}
\end{table}

\subsection{Overall Performance Comparison}

In this comprehensive experimental study, we evaluated the performance of several state-of-the-art recommendation methods across diverse datasets: Beauty and Toys-and-Games, using Recall@20, Recall@50, NDCG@20, and NDCG@50. Predominantly, SGCL achieved the highest NDCG and Recall scores across all datasets, highlighting its superior efficacy in recommendation tasks. This outcome not only emphasizes the significance of the augmentation-free framework but also validates the effectiveness of the supervised graph contrastive loss optimization.
Most of the self-supervised graph recommender systems consistently outperform the traditional ones. This suggests that the auxiliary contrastive learning task leverages extra graph structure information, boosting the performance of the predictions on unobserved user-item interactions.

\subsection{Efficiency Analysis}
\subsubsection{\textbf{Trade-off between the Performance and the efficiency.}}

\begin{figure}[htpb]
    \centering
    \includegraphics[width=0.75\linewidth]{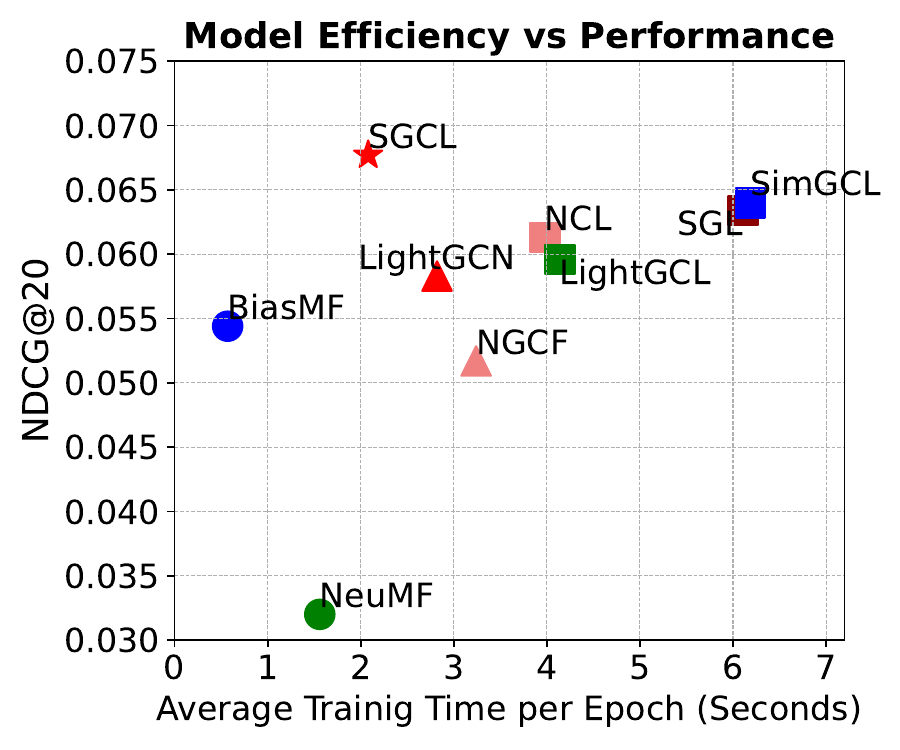}
    \caption{Trade-off between the performance and the efficiency on the Beauty dataset. The upper side indicates better performance; the left side represents more efficient training.
    }
    \label{fig: tradeoff}
\end{figure}

Figure \ref{fig: tradeoff} demonstrates the superiority of our models in both efficiency and performance. Two matrix factorization models are more efficient than graph models but achieve much lower performance. LightGCN and NGCF outperform the rest of the graph contrastive learning baselines in terms of speed. Notably, our method SGCL exhibits the most competitive performance while being much faster than the other graph contrastive recommendation baselines. 

\subsubsection{\textbf{Training Time Comparison.}}
In Table~\ref{tab: time}, LightGCN is efficient per epoch but may not always be the best choice when considering total training time due to the higher number of epochs required. SGL appears to offer a better compromise between the number of epochs and time per epoch, often resulting in an approximate total time compared with LightGCN. 
In contrast, SGCL demonstrates extreme efficiency in all datasets, which run for the fewest epochs despite being the fastest per epoch in most cases, thereby resulting in the lowest total training time in all datasets.

\begin{table}[htpb] 
\begin{center}
\small
\caption{Training time efficiency comparisons (we denote second as s, and minute as m as the abbreviation).}
\label{tab: time}
  \begin{tabular}{ccccc}
    \toprule

Dataset  & Method  & Time/Epoch & \# Epochs & Total Time \\
\midrule
\multirow{4}{*}{Beauty} 
&LightGCN    &2.82s	&153&	7.19m \\
&SGL         &6.11s	&63&	6.42m \\
&SGCL        &2.08s	&62&	2.15m \\
\midrule
\multirow{4}{*}{\makecell[l]{Toys-and\\-Games}} 
&LightGCN    &1.53s	&173&	4.41m \\
&SGL         &6.24s	&75&	7.80m \\
&SGCL        &1.63s	&35&	0.95m \\

  \bottomrule
\end{tabular}
\end{center}
\end{table}

\subsubsection{\textbf{Performance Curve and Convergence Speed}}
\begin{figure}[htpb]
    \centering
    \begin{subfigure}[b]{0.95\linewidth}
        \centering
        \includegraphics[width=\textwidth]{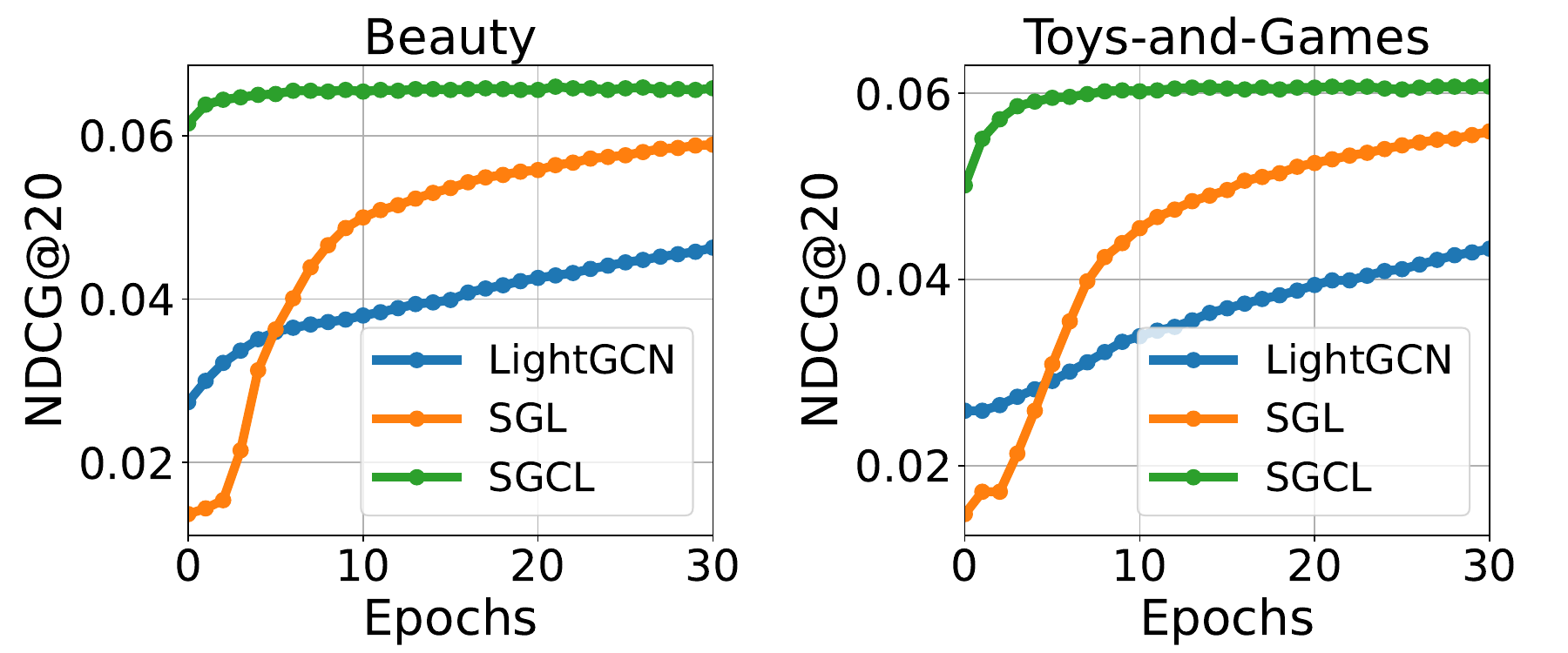} 
    \end{subfigure}
    \caption{Performance of NDCG in the first 30 epochs.}
    \label{fig: curve}
\end{figure}

We compare the NDCG@20 (Figure \ref{fig: curve}) across the first 30 training epochs with the LightGCN and the SGL on three datasets. By 30 epochs, both of our model SGCL achieve high and stable Recall levels, whereas the LightGCN and the SGL have been far from convergence points. In consonance with the outcomes of our training time experiments in Figure~\ref{fig: tradeoff}, SGCL demonstrates low total training time and fast convergence speed. These findings underscore the preeminence of SGCL in practical and real-world applications.

\section{Related Work}

Recently, self-supervised contrastive learning has gained considerable attention across various fields due to its proficiency in handling massive amounts of unlabeled data for the data sparsity issue \cite{chen2020simple, tian2020makes, you2020graph, peng2020graph, hassani2020contrastive}. 
Motivated by this, NCL\cite{lin2022improving} incorporates potential neighbors into contrastive pairs, by considering both the graph structure and semantic space. Differently, the dropout-based methods are employed by SGL\cite{wu2021self} for randomized graph augmentation to obtain two contrastive views. The optimization includes node-level contrast using the InfoNCE loss, along with the BPR loss for recommendation. SimGCL\cite{yu2022graph} first investigates the necessity of complex graph augmentation and then devises noise-based augmentation techniques that perturb the representations of users and items during training. Then, LightGCL \cite{cai2023lightgcl} has been proposed to utilize the singular value decomposition (SVD) \cite{rajwade2012image} as a light tool to augment user-item graph structures. Nevertheless, all the previous works follow the multi-task learning paradigm with tedious manual tuning of task weight and require additional graph encoders as well as different augmentation procedures during training.

\section{Conclusion}

In this paper, we revisit the existing graph contrastive learning methodologies in the recommendation and identify \textbf{\textit{Ineffective Optimization}} challenge. Towards this issue, SGCL innovatively integrates the training of recommendation and contrastive losses using a supervised contrastive learning loss. This unified approach effectively resolves the problem of inconsistent optimization directions and the inefficient graph convolutions, typically seen in decoupled multi-task learning frameworks. The empirical validation of SGCL, conducted through extensive experiments on real-world datasets, underscores its superior performance. Notably, SGCL outshines state-of-the-art methods, achieving faster convergence speed, and excelling in both accuracy and efficiency. This paper's findings highlight the potential of coupled design in supervised graph contrastive learning, paving the way for more efficient and effective recommender systems.

\begin{acks}
This work is supported in part by NSF under grants III-2106758, and POSE-2346158.
\end{acks}
\bibliographystyle{ACM-Reference-Format}
\bibliography{sample-base}



\end{document}